%% file: main.tex
\documentclass{article}
\usepackage[T1]{fontenc}
\usepackage{microtype}
\usepackage{graphicx}
\usepackage{subcaption}
\usepackage{booktabs} 
\usepackage{tabularx}
\usepackage{hyperref}
\usepackage{xspace}
\newcommand{\eg}{e.g.\xspace}
\usepackage[table]{xcolor}

\usepackage{tech2025}

\usepackage{amsmath}
\usepackage{amssymb}
\usepackage{mathtools}
\usepackage{amsthm}
\usepackage{tcolorbox}

\usepackage[capitalize,noabbrev]{cleveref}
\usepackage{fontawesome}

\definecolor{bigaired}{RGB}{156, 0, 0}
\definecolor{uclablue}{RGB}{39, 116, 174}

\definecolor{darkred}{RGB}{200, 0, 0}
\definecolor{darkblue}{RGB}{0, 0, 200}
\definecolor{blue}{RGB}{0, 0, 250}

\definecolor{light}{RGB}{225, 250, 250}
\definecolor{lightgray}{RGB}{0.9, 0.9, 0.9}
\definecolor{lightred}{RGB}{250, 200, 200}
\definecolor{lightblue}{RGB}{210, 220, 250}

\definecolor{doderblue}{RGB}{30, 144, 255}
\definecolor{select}{RGB}{222, 235, 247}
\definecolor{unselect}{RGB}{247, 207, 206}

\definecolor{lightgrey}{RGB}{247, 247, 247}

\hypersetup{colorlinks=true, citecolor=uclablue, linkcolor=uclablue, urlcolor=bigaired}

\newenvironment{bigaiabstract}{
  \begin{tcolorbox}[
    colback=lightgrey,
    colframe=white,
    boxrule=0pt,
    arc=10pt,
    left=16pt,
    right=16pt,
    top=12pt,
    bottom=12pt,
    width=\textwidth,
    enlarge left by=0mm,
    before skip=10pt,
    after skip=10pt
  ]
  \normalsize
}{
  \end{tcolorbox}
}

\makeatletter
\fancypagestyle{fancytitlepage}{
  \fancyhead{}  
  \lhead{\includegraphics[height=1.5cm]{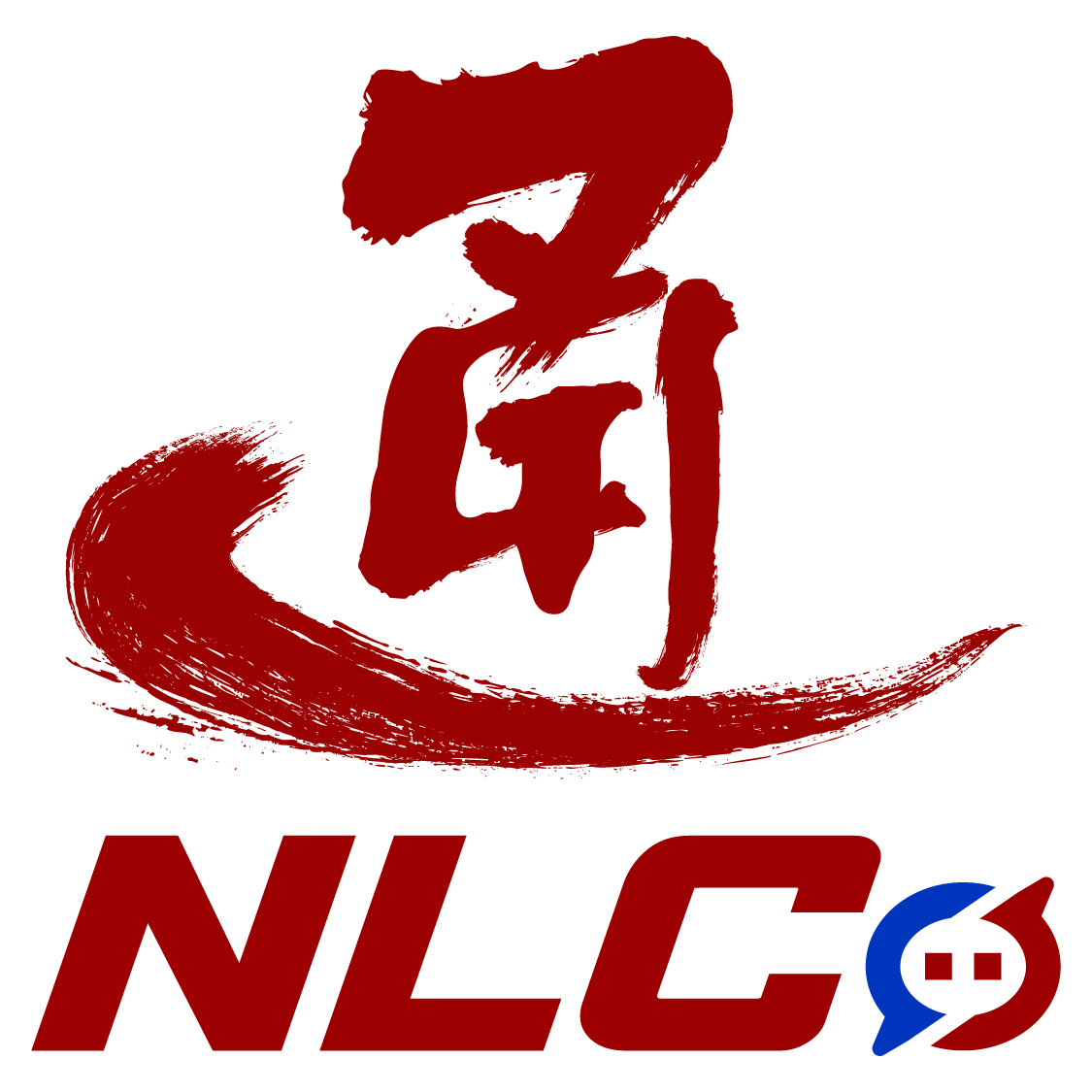}}
  \rhead{\it \@icmldate}
  \cfoot{}
}
\makeatother

\theoremstyle{plain}

\theoremstyle{definition}

\theoremstyle{remark}

\usepackage[textsize=tiny]{todonotes}

\newcommand{\methodlong}{TongSearch QR}
\newcommand{\ourmethod}{TongSearch-QR}
\begin{document}

\icmldate{\today}

\icmltitle{TongSearch-QR: Reinforced Query Reasoning for Retrieval}



\begin{icmlauthorlist}
\icmlauthor{Xubo Qin}{bigai}
\icmlauthor{Jun Bai}{bigai}
\icmlauthor{Jiaqi Li}{bigai}
\icmlauthor{Zixia Jia}{bigai}
\icmlauthor{Zilong Zheng}{bigai}
\end{icmlauthorlist}

$^{1\,}$NLCo Lab, Beijing Institute for General Artificial Intelligence (BIGAI)
\icmlaffiliation{bigai}{}

\icmlcorrespondingauthor{Zilong Zheng}{zlzheng@bigai.ai}
\icmlcorrespondingauthor{Zixia Jia}{jiazixia@bigai.ai}
\icmlkeywords{Machine Learning, ICML}

\vskip 0.3in



\printAffiliationsAndNotice{}  

\begin{bigaiabstract}
Traditional information retrieval (IR) methods excel at textual and semantic matching but struggle in reasoning-intensive retrieval tasks that require multi-hop inference or complex semantic understanding between queries and documents.
One promising solution is to explicitly rewrite or augment queries using large language models (LLMs) to elicit reasoning-relevant content prior to retrieval. However, the widespread use of large-scale language models like GPT-4 or LLaMA3-70B remains impractical due to their high inference cost and limited deployability in real-world systems.
In this work, we introduce \textbf{\ourmethod}, a family of small-scale language models for query reasoning and rewriting in reasoning-intensive retrieval. With a novel semi-rule-based reward function, we employ reinforcement learning approaches enabling smaller language models, \eg, \texttt{Qwen2.5-7B-Instruct} and \texttt{Qwen2.5-1.5B-Instruct}, to achieve query reasoning performance rivaling large-scale language models without their prohibitive inference costs. Experiment results on BRIGHT~\cite{su2024bright} benchmark show that with BM25 as retrievers, both \texttt{\ourmethod-7B} and \texttt{\ourmethod-1.5B} models significantly outperform existing baselines, including prompt-based query reasoners and some latest dense retrievers trained for reasoning-intensive retrieval tasks, offering superior adaptability for real-world deployment. 

\vskip .1in

\begin{center}
\setlength{\tabcolsep}{5pt}
    \begin{tabular}{ccc}
        {\large \faGithub} & \textbf{Code} & \url{https://github.com/bigai-nlco/TongSearch-QR} \\
    \end{tabular}
\end{center}

\end{bigaiabstract}

\vskip .3in

\section{Introduction} 

\begin{figure}
\centering
\includegraphics[scale=0.9]{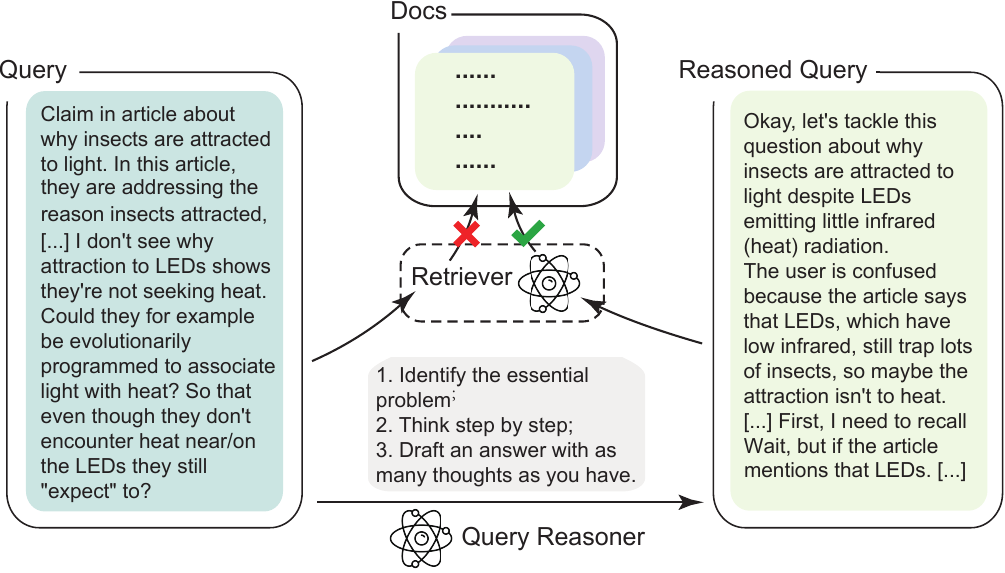}
\caption{
An example query reasoning with LLM. The query is sampled from the Biology Subtask of BRIGHT~\cite{su2024bright} benchmark.}
\label{query_reasoning_full}
\end{figure}
The Information retrieval system (IR)~\cite{zhu2023large} plays a critical role in enabling users to locate relevant materials from vast repositories of documents, Web pages, and structured records~\cite{bajaj2018msmarcohumangenerated,denseretrieval_qa,imageretrieval}. Existing retrieval methods mainly focus on measuring the relevance between queries and documents via text matching or semantic representation techniques, e.g., BM25 algorighm~\cite{bm25} or document embedding models~\cite{devlin2018bert,liu2019roberta,chen_bge_2024,ma_fine-tuning_2024}, achieving considerable success~\cite{liu2021retrieval_baidu,jdrel,web_search_tencent}. However, in real-world applications, users may issue questions with high complexity, and finding the relevant documents requires intensive reasoning~\cite{su2024bright}. For example, a programmer may ask a question to find a function (denoted as $Func_b$) that can be an alternative option to the given function $Func_a$. The ground truth document may be an introduction of $Func_b$, while it may be completely disconnected from the question, both lexically and semantically. The document might not mention $Func_a$ at all, nor contain any explicit cues suggesting its substitutability. In such cases, traditional retrieval methods that rely on lexical overlap or shallow semantic similarity often fail, as they cannot capture the implicit reasoning chain required to connect the user’s intent (i.e., “find an alternative to $Func_a$”) to the actual answer (i.e., “The description of $Func_b$ ”).

This highlights a fundamental challenge in information retrieval for complex queries: the need to bridge a reasoning gap between the user’s implicit intent and the relevant knowledge. An effective retrieval system must go beyond matching surface-level expressions—it needs to infer what the user is fundamentally trying to achieve, and then identify which pieces of information in the corpus instantiate or fulfill that abstract goal. This requires not only modeling the latent intent behind a query, but also mapping that intent to the appropriate segments of knowledge, which may be distributed, implicit, or expressed in entirely different terms. These kinds of retrieval tasks demand models capable of aligning abstract user goals with semantically distant but conceptually relevant content. We refer to such tasks as \textbf{reasoning-intensive retrieval}~\cite{su2024bright,shao2025reasonir}, which has been proved to be challenging for most of the existing retrieval approaches with poor performance.

To address this issue, two research directions have been proposed. One is to train novel retriever or reranker models~\cite{shao2025reasonir,weller2025rank1} with task-specific reasoning data. The other is to apply \textbf{query reasoning and rewriting} to the given query~\cite{su2024bright,niu2024judgerank,jagerman_query_2023}, leveraging the frontier reasoning capabilities of large language models (LLMs)~\cite{guo2025deepseek,openr1} with chain-of-thought reasoning~\cite{wei2022chain} to generate an intermediate reasoning result as \textbf{reasoned query}, which instead will be used to retrieve the relevant documents. Figure~\ref{query_reasoning_full} shows an example of query reasoning. \textbf{These two directions are orthogonal}: a retriever designed for reasoning-intensive tasks can take the reasoned queries as input, leading to further improvements of retrieval performance. Existing query reasoning approaches mainly rely on large-scale LLMs (\eg, GPT-4o~\cite{openai2024gpt4ocard} or LLama3-70B~\cite{grattafiori2024llama3herdmodels}) with chain-of-thought prompts. This limits the applicability of such methods, as in many real-world RAG scenarios, high-performance commercial models like GPT-4o are not accessible due to inference cost or information security concerns. Moreover, the query rewriting model deployed is generally not expected to outperform the main model~\cite{shao2025reasonir} in RAG pipelines. For example, suppose that an RAG system is primarily built around a small-scale model instance such as Qwen2.5-7B-Instruct. In that case, the query reasoning module is typically constrained to use a model of equal or smaller capacity, which cannot outperform the main model of Qwen2.5-7B mentioned above. As a result, query reasoning methods that rely on high-performance models and Chain-of-Thought prompts are not feasible in such realistic RAG settings.

\begin{figure}[!ht]
    \centering
    \begin{minipage}[c]{0.58\linewidth}  
        \centering
        \includegraphics[width=\linewidth]{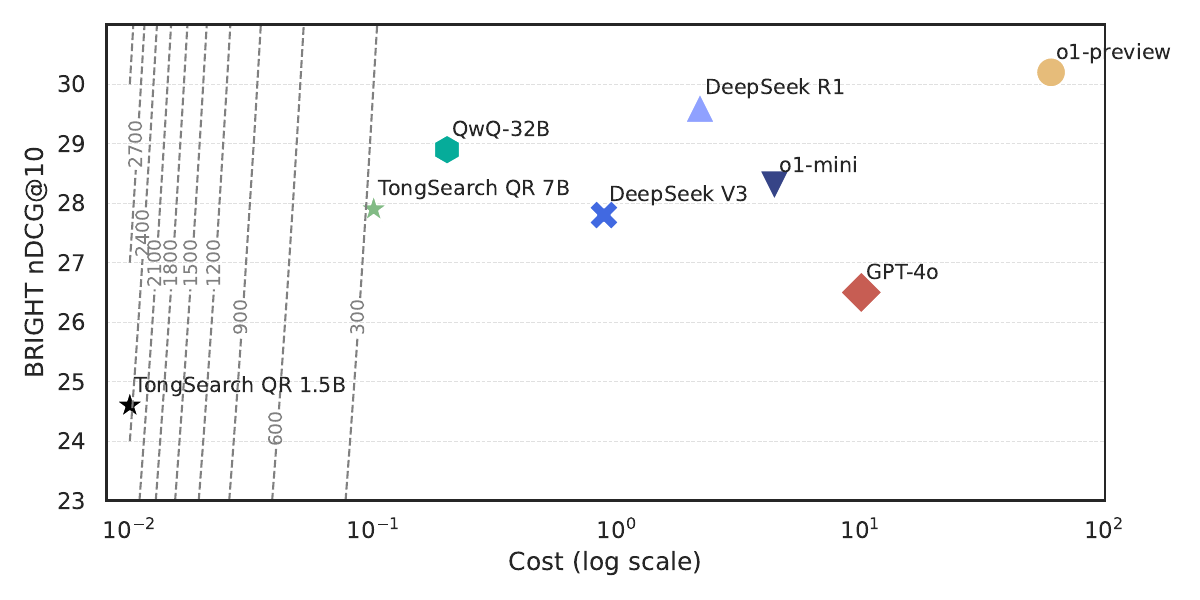}
        \captionsetup{width=0.9\linewidth}
        \caption{Cost vs. Performance comparison of different models. Details about the cost and performance can be found in Table~\ref{tab:model_efficiency}. }
        \label{fig:task_mixing_recipe}
    \end{minipage}
    \hfill 
    \begin{minipage}[c]{0.39\linewidth}  
        \centering
        \includegraphics[width=\linewidth]{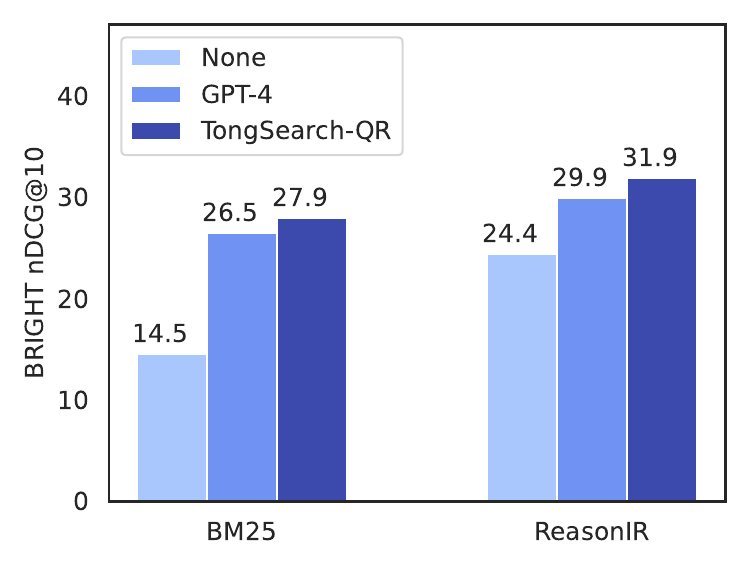}
        \captionsetup{width=0.9\linewidth}
        \caption{Performance on different reasoning models. Details can be found in Table~\ref{tab: bright_result}}
        \label{fig:sample_efficiency}
    \end{minipage}
\end{figure}

In this paper, we introduce \ourmethod, a family of small-scale language models for query reasoning and rewriting (QR). To the best of our knowledge, this is the first model family specifically trained for query reasoning in reasoning-intensive retrieval tasks. Inspired by previous works using Reinforcement Learning with Verifiable Rewards (RLVR) to enhance LLMs' reasoning~\cite{guo2025deepseek,qwen2025qwen25technicalreport}, we developed a novel semi-rule-based reward function for GRPO~(Group Relative Policy Optimization)~\cite{shao2024deepseekmath,guo2025deepseek}, enabling RL on the query reasoning of smaller language models. Beyond that, we propose an automatic data curation pipeline for training reasoning-based rewriting with publicly available dataset~\cite{h4stackexchange}. Experiment results on the BRIGHT~\cite{su2024bright} benchmark show that our model achieves an NDCG@10 metric of $27.9$, outperforming GPT-4o's metric of $26.5$. This metric is comparable to some large-scale reasoning models, e.g., o1-preview\footnote{https://openai.com/index/introducing-openai-o1-preview/}, DeepSeek R1\cite{guo2025deepseek}, and QwQ-32B~\cite{qwen2025qwen25technicalreport}, with significantly lower cost of inference shows in Figure~\ref{fig:task_mixing_recipe}. 
Besides, our proposed models can also work with reasoning-intensive retrievers~\cite{shao2025reasonir} to achieve the best performance, as shown in Figure~\ref{fig:sample_efficiency}. This demonstrates that our models possess strong flexibility to adapt to different retrieval pipelines. 

In summary, our \textbf{main contributions} are listed as follows:
\begin{itemize}
    \item \textbf{Query reasoning models for retrieval tasks:} We propose \ourmethod family (7B and 1.5B) specifically trained for query reasoning and rewriting in reasoning-intensive retrieval tasks. Our small-scale language models are comparable to state-of-art large-scale language models such as GPT-4o on specific tasks. These results make it possible to apply query reasoning in many real-world RAG system settings. Besides, our query reasoning models can be jointly applied to different existing retrievers to achieve better performance.  
    \item \textbf{Semi-rule-based reward function for RL:} The reward function inherits the advantage of existing functions based on semantic similarity, which evaluates the relevance enhancement between queries and retrieval documents. It offers a range of advantages, including strong robustness, high computational efficiency, and the avoidance of reward hacking. 
    \item \textbf{Automatic data curation pipeline:}  The data curation pipeline proposed in this paper is specifically designed to build training data for query rewriting tasks. It optimizes training without the need for large-scale supervised query reasoning data, which is often unavailable in real applications and scenarios.
\end{itemize}

\section{Related Works}
\paragraph{Reasoning-intensive Retrieval}
In recent years, dense retrieval has achieved remarkable progress in retrieval accuracy, propelled by the rapid evolution of foundation models and innovative training methodologies~\cite{luo2024large,gao2022precise,lee2024nv,wei2024uniir}. Nowadays, BERT~\cite{devlin2018bert}-based and LLM-based~\cite{wang2023improving,luo2024large} embedding models have been widely used in multiple retrieval tasks, achieving great success as general-purpose retrievers~\cite{wang2022text,li2023towards,chen_bge_2024,khattab2020colbert}. However, recent work of BRIGHT Benchmark~\cite{su2024bright} has demonstrated that most of the existing BERT-based or LLM-based retrievers and re-rankers perform poorly on the task of reasoning-intensive retrieval~\cite{shao2025reasonir}. These results suggest that existing retrieval methods, which assess relevance based on textual or semantic similarity, fall short in capturing the deep-dived, reasoning-driven relevance that arises in complex knowledge-seeking scenarios. To address this issue, some researchers tried to train reasoning-enhanced retrievers~\cite{shao2025reasonir} or rerankers~\cite{weller2025rank1} with public or LLM-Synthesized datasets. Another way is to apply LLMs for query reasoning and rewriting. The LLMs take the original queries as input to generate Chain-of-Thought reasoning steps as pseudo queries. These pseudo queries contain richer knowledge and contextual information related to the original question, they are more likely to retrieve relevant documents than the original queries when issued to a text-based or semantic-based retriever. The two approaches mentioned above are orthogonal and can be combined synergistically. Most of those existing query reasoning approaches~\cite{su2024bright,niu2024judgerank} are based on prompting large-scale LLMs, \eg, GPT-4o~\cite{openai2024gpt4ocard} or LLama3-70B~\cite{grattafiori2024llama3herdmodels}, which are not available in many real-world settings of RAG systems. To the best of our knowledge, none of those previous works focuses on training a small-scale language model for query reasoning and rewriting tasks.

\paragraph{Reinforcement Learning with Verifiable Reward}
Large reasoning models, \eg, \texttt{OpenAI o1}, \texttt{Gemini Flash-Thinking}\footnote{https://deepmind.google/technologies/gemini/flash-thinking/}, \texttt{DeepSeek-R1}~\cite{guo2025deepseek} and \texttt{QwQ-32B}~\cite{qwen2025qwen25technicalreport}, have achieved great success in reasoning-intensive areas like coding and mathematical proofs. These models adopt a ``slow-thinking''~\cite{wu2024inference,chen2024llmcallsneedscaling} approach for the models to first output a sequence of thinking processes with the tags of ``<think></think>'' before providing the actual answer. This method has allowed LLMs to enhance their reasoning capabilities to achieve further performance improvements on math or coding tasks~\cite{openr1}. Inspired by the technical report released by DeepSeek~\cite{guo2025deepseek}, researchers~\cite{openr1,xie2025logicrl} have tried to reproduce the slow-thinking ability on smaller-scaled LLMs via reinforcement learning based on GRPO (Group Relative Policy Optimization)~\cite{shao2024deepseekmath} and rule-based reward functions. Compared with model-based reward functions (e.g., reward functions based on process reward model~\cite{prmlessons}), verifiable reward functions have the advantages of being simple and effective, making the model training process easier to scale up. Besides, rule-based reward functions focus solely on the correctness of output results, ignoring the intermediate process, which makes them immune to reward hacking and increases the robustness of model training.  Moreover, unlike supervised fine-tuning (SFT), reinforcement learning based methods do not force the model to fit every generated token, thereby yielding superior generalization capabilities.

\section{\methodlong}
\subsection{Task Formulation}
Given a query $q$ and a set of candidate documents $D$ = $\{d_1, ..., d_n\}$, the objective of information retrieval task is to identify and retrieve a subset of relevant documents from $D$: $D^+$ = $\{d^+_1, ...,d^+_i,..., d^+_m\}$, where $m\ll n$ leveraging a retriever $\mathcal{RT}$. 
In reasoning-intensive retrieval, instead of text or semantic similarity,  \( Q \) is relevant to \( D^+ \) through a specific reasoning path or explanation (e.g., underlying principles, algorithms, or theorems) associated with the query. For instance, typical reasoning paths may involve identifying the query intent, analyzing and modeling the problem, and deriving sub-conclusions based on the provided descriptions. Such reasoning paths usually do not exist explicitly in the query itself, making direct retrieval based solely on the query highly challenging. In this paper, we refer to the process of generating reasoning paths or explanations for query \( Q \) as \textbf{query reasoning}.  

In this paper, we denote $\mathcal{LLM}$ as a large language model for query reasoning and generating the rewritten query $q^{'}$ based on $q$.  $\mathcal{RT}$ will later use $q^{'}$ to retrieve the documents relevant to $q$. The processes mentioned above can be described with the following equations:
\begin{align*}
    q^{'}&=\mathcal{LLM}(\text{Inst};q), & D^+ =\mathcal{RT}(q^{'}),
\end{align*}
where $\text{Inst}$ denotes the instructions for query reasoning and rewriting. 

\subsection{Reinforcement Learning with Semi-Rule-Based Reward} 
\label{semi-rule-reward}
\paragraph{Preliminary} Inspired by previous works of large reasoning models \eg, DeepSeek R1~\cite{guo2025deepseek}, we employ the GRPO-based reinforcement learning algorithm to train the LLMs for query reasoning, where the model takes the given query $q$ as input and generates a reasoned query $q^{'}$. The GRPO objective is defined as:

\begin{equation*}
\begin{aligned}
\mathcal{L}_{\text{GRPO}}(\theta) &= \mathbb{E}_{(q,a) \sim \pi_\theta} \big[ w_g \cdot 
\min\big( r_\theta(q, q^{'}) \cdot \hat{A}(q, q^{'}), \\
        & \text{clip}(r_\theta(q, q^{'}), 1 - \epsilon, 1 + \epsilon) \cdot \hat{A}(q, q^{'}) \big) \big]
\end{aligned}
\end{equation*}

Here, $r_\theta(q, q^{'}) = \frac{\pi_\theta(q^{'}|q)}{\pi_{\theta_{\text{old}}}(q^{'}|q)}$ is the importance ratio between the current and reference policy. The advantage function $\hat{A}(q, q^{'})$ is computed based on the group-normalized reward:

\begin{equation*}
\hat{A}(q, q^{'}) = \frac{R(q, q^{'}) - \mu_g}{\sigma_g + \delta},
\end{equation*}
where $R(q, q^{'})$ is the reward assigned to the reasoned query $q^{'}$, $\mu_g$ and $\sigma_g$ denote the mean and standard deviation of rewards within the group $g$, and $\delta$ is a small constant to avoid division by zero. The weight $w_g$ optionally rescales the advantage based on group-level reward variance. This formulation stabilizes training when rewards are sparse or highly variable across different query groups.

\paragraph{Limitations for Previous Rule-based Reward Function} Previous approaches~\cite{jiang2025deepretrieval} of rule-based reward for retrieval tasks are usually calculated based on retrieval evaluation metrics like Recall@K. The metric-based reward function requires both annotated training data and an existing large-scale document collection to serve as the retrieval source, which is difficult to access in reasoning-intensive retrieval tasks.

\paragraph{Semi-Rule-Based Reward for Query Reasoning} In this work, we introduce a reward function to evaluate the incremental relevance score from $<q, D^{+}>$ to $<q^{'}, D^{+}>$. For a reasoning-intensive task, the goal of query reasoning and rewriting is to improve the retrieval performance using a reasoned query $q^{'}$ with a higher relevance score compared to $q$. Since the relevance score is computed via an existing relevance model, the reward function is defined as a ``semi-rule-based reward function''.

Each training sample consists of $<q, D^{+}>$, where $D^{+}$ indicates single or multiple positive documents for $q$. We define $score_{q}$ as the sum of the relevance scores between $q$ and each positive document in $D^{+}$:
\begin{align*}
    score_{q}=\Sigma_{i\in D^{+}}\text{Rel}(q,d^{+}_{i})
\end{align*}
where $Rel(q,d^{+}_{i})$ denotes the relevance score between $q$ and $d^{+}_{i}$ computed via a relevance model. Here we use a pretrained embedding model to encode queries and documents into embeddings, with the cosine similarities as relevance scores. The parameters of the relevance model will not be updated during the model training process.
Similarly, the score of the reasoned query $score_{q^{'}}$ is also computed as:
\begin{align*}
    score_{q^{'}}=\Sigma_{i\in D^{+}}\text{Rel}(q^{'},d^{+}_{i})
\end{align*}
The overall reward is defined as the average relevance score increment from $q$ to $q^{'}$ of each positive document:
\begin{align*}
    R(q, q^{'})=\frac{score_{q^{'}}-score_{q}}{|D^{+}|}
\end{align*}


Our semi-rule-based reward function inherits a few advantages from the existing rule-based rewards as follows: 
Firstly, the function depicts the semantic relevance based on the existing embedding model like bge-base-en~\cite{chen_bge_2024}, which has been proved to exhibit good performance with robustness and low computational cost.
Secondly, unlike the process reward models (PRMs), our method
does not rely on intermediate processes supervision, and is therefore inherently immune to reward hacking. These properties collectively contribute to the high computational efficiency and robustness of our method, enhancing its tolerance to noise present in the training data.

\subsection{Training Data Curation}
\label{dataset_construct}
Existing training datasets like \eg, MSMACRO~\cite{bajaj2018msmarcohumangenerated} are helpful for semantic-based retrieval tasks, which are not specifically designed for reasoning-intensive retrieval. 
Inspired by the data construction process in benchmark BRIGHT~\cite{su2024bright}, we use the publicly available H4 Stack Exchange Preferences~\cite{h4stackexchange} dataset to construct our training data. The dataset contains questions and answers from the Stack Overflow Data Dump for the purpose of preference model training. Each question in the dataset includes at least two answers, and each answer is labeled ``is\_selected'' or not, indicating if the answer is selected and marked as useful by the real users who issued the question. We select QAs with texts only for data curation.

Here are two ways we further obtain the rewritten queries as the ``supervision'' for query reasoning training: 

(1) Given a query for reasoning, a large reasoning model, \eg, QwQ-32B or DeepSeek-R1, is asked to generate the rewritten query based on Chain-of-Thought(CoT) reasoning. The curated data is denoted as \textbf{V1-R1} and \textbf{V1-QwQ}. 

(2) For each question, we use the answer with the ``selected'' tag as the reasoned query from StackExchange by real users, which is denoted as \textbf{V2}. Notice that not every question includes a selected answer.

Details about the actual dataset in use will be further described in Section~\ref{dataset} and Appendix~\ref{app:training_data_detail}.

\begin{table*}[h]
    \centering
    \resizebox{\textwidth}{!}{
    \begin{tabular}{l|cccccccccccc|c}
        \toprule
        & \multicolumn{7}{c}{\textbf{StackExchange}} & \multicolumn{2}{c}{\textbf{Coding}} & \multicolumn{3}{c}{\textbf{Theorem-based}} & \multicolumn{1}{c}{\textbf{Avg}} \\
        \cmidrule(lr){2-8} \cmidrule(lr){9-10} \cmidrule(lr){11-13} \cmidrule(lr){14-14}
        & Bio. & Earth. & Econ. & Psy. & Rob. & Stack. & Sus. & Leet. & Pony & AoPS & TheoQ. & TheoT. & \\
        \midrule
        \multicolumn{14}{c}{\textit{\textbf{\cellcolor{lightgray!40}Retrievers with Original Queries}}} \\
        \hline
        BM25 & 18.9 & 27.2 & 14.9 & 12.5 & 13.6 & 18.4 & 15.0 & 24.4 & 7.9 & 6.2 & 10.4 & 4.9 & 14.5 \\
        BGE & 11.7 & 24.6 & 16.6 & 17.5 & 11.7 & 10.8 & 13.3 & 26.7 & 5.7 & 6.0 & 13.0 & 6.9 & 13.7 \\
        ReasonIR & 26.2 & 31.4 & 23.3 & 30.0 & 18.0 & 23.9 & 20.5 & 35.0 & 10.5 & 14.7 & 31.9 & 27.2 & 24.4 \\
        Seed1.5-Embedding & 34.8 & 46.9 & 23.4 & 31.6 & 19.1 & 25.4 & 21.0 & 43.2 & 4.9 & 12.2 & 33.3 & 30.5 & 27.2  \\
        \hline
         \multicolumn{14}{c}{\textit{\textbf{\cellcolor{lightgray!40}Query Reasoner with BM25}}} \\
        \hline
        GPT-4o & 53.6 & 53.6 & 24.3 & 38.6 & 18.8 & 22.7 & 25.9 & 19.3 & 17.7 & 3.9 & 18.9 & 20.2 & 26.5 \\
        Doubao & 54.8 & 53.3 & 23.7 & 37.2 & 22.2 & 28.1 & 25.0 & 21.2 & 16.4 & 7.8 & 21.8 & 22.7 & 27.8 \\
        Deepseek-V3 & 56.6 & 54.2 & 25.8 & 38.8 & 19.9 & 26.7 & 26.4 & 19.8 & 15.1 & 6.7 & 22.5 & 20.7 & 27.8 \\\hline
        o1-mini & 60.2 & 57.4 & 24.7 & 39.3 & 23.3 & 26.4 & 25.4 & 23.5 & 13.4 & 6.9 & 22.8 & 16.5 & 28.3 \\
        o1-preview & \textbf{64.2} & \underline{57.9} & 27.6 & \textbf{43.1} & \textbf{25.6} & 29.1 & 28.0 & 21.2 & 15.9 & 5.6 & 24.0 & 20.5 & 30.2 \\
        Deepseek-R1 & \underline{62.7} & \textbf{58.3} & 26.0 & \underline{42.9} & 21.8 & 28.1 & \textbf{30.3} & 19.6 & 10.7 & 6.0 & 25.8 & 22.4 & 29.6 \\
        R1-distill-qwen-7B & 33.9 & 41.6 & 19.9 & 31.8 & 15.1 & 18.8 & 16.4 & 19.7 & 10.7 & 6.8 & 24.5 & 22.2 & 21.8 \\
        R1-distill-qwen-32B & 50.6 & 49.9 & 22.9 & 38.1 & 20.3 & 24.6 & 19.2 & 19.5 & 11.3 & 5.6 & 24.2 & 20.2 & 25.5 \\
        QwQ-32B & 57.5 & 56.3 & 29.9 & 41.8 & 19.2 & 25.7 & 27.2 & 21.5 & 12.8 & 6.5 & 25.4 & 22.8 & 28.9 \\
        \midrule
       
        \textbf{\ourmethod-1.5B} & 46.0 & 47.1 & 21.1 & 31.2 & 19.8 &21.7 & 24.3 & 22.5 & \textbf{21.7} & 4.3 & 19.7 & 15.9 & 24.6 \\
         \textbf{\ourmethod-7B} & 57.9 & 50.9 & 21.9 & 37.0 & 21.3 & 27.0 & 25.6 & 23.6 & 14.4 & 7.0 & 26.1 & 22.0 & 27.9 \\
        \midrule
        \multicolumn{14}{c}{\textit{\textbf{\cellcolor{lightgray!40} Query Reasoner with ReasonIR}}} \\
        \hline
        LLama3.1-8B-Instruct  & 37.8 & 39.6 & 29.6 & 35.3 & 24.1 & \textbf{31.1} & 27.4 & 28.8 & 14.5 & 9.2 & 26.6 & 32.3 & 28.0 \\
        GPT-4  & 43.6 & 42.9 & \textbf{32.7} & 38.8 & 20.9 & 25.8 & 27.5 & \underline{31.5} & \underline{19.6} & 7.4 & 33.1 & \underline{35.7} & \underline{29.9} \\
        \textbf{\ourmethod-1.5B}  & 36.4 & 41.1 & 29.9 & 34.0 & 25.2 & \underline{30.7} & 25.6 & \textbf{33.3} & 16.8 & \underline{9.7} & \underline{35.7} & 32.7 & 29.3 \\
        \textbf{\ourmethod-7B}  & 46.2 & 45.1 & \underline{31.2} & 39.6 & \underline{25.3} & 28.7 & \underline{28.4} & 31.2 & 16.3 & \textbf{10.8} & \textbf{40.0} & \textbf{39.3} & \textbf{31.9} \\
        \bottomrule
    \end{tabular}}
    \caption{Performance comparison on BRIGHT. The best score is shown in bold and the second best is underlined.}
    \label{tab: bright_result}
\end{table*}
\section{Experiment}
\subsection{Experimental Setup} 
\subsubsection{Dataset and Metrics}
\label{dataset}
\paragraph{Training} 
We employ two types of the constructed data mentioned in Section~\ref{dataset_construct} for training: 
\textbf{V1-R1}, \textbf{V1-QwQ} and \textbf{V2}. For V2, we use the user-selected answers since the size of V2 is too large to afford the inference cost of large reasoning models. More details can be found in Appendix~\ref{app:training_data_detail}.

\paragraph{Evaluation} We use BRIGHT~\cite{su2024bright}, a novel benchmark for reasoning-intensive retrieval that aims to evaluate the ability of retrieval models to handle complex queries that require deep reasoning. It consists of 1,384 real-world queries from diverse domains with 12 sub-tasks. We adopt the  metric \textbf{nDCG@10} for the following evaluations.

\subsubsection{Baselines}
The baselines in our experiments can be divided into these three categories:

\paragraph{Retrievers with Original Queries} There are two types of baselines: 1) Traditional baselines in IR systems like BM25~\cite{bm25} for sparse retrieval and bge-large-en~\cite{chen_bge_2024}for dense retrieval; 2) Reasoning-intensive retrievers like ReasonIR~\cite{shao2025reasonir} and Seed 1.5-Embedding\footnote{https://huggingface.co/ByteDance-Seed/Seed1.5-Embedding}.  We keep the same with the experiments reported in~\cite{su2024bright} for fair comparison and all the retrievers use the original queries in BRIGHT to retrieve documents. Since Seed1.5-Embedding is not public available when this work is done, we directly use the experiment results reported on their model card.

\paragraph{Query Reasoner with BM25}  
We include two types of baselines using state-of-the-art large language models: 1) Non-reasoning models including GPT-4o, doubao-1.5-pro~\footnote{\text{https://seed.bytedance.com/en/special/doubao\_1\_5\_pro}}, DeepSeek-V3~\cite{deepseekai2025deepseekv3technicalreport}; 2) Reasoning models including DeepSeek R1~\cite{guo2025deepseek}, o1-mini\footnote{https://openai.com/index/openai-o1-mini-advancing-cost-efficient-reasoning/}, o1-preview\footnote{https://openai.com/index/introducing-openai-o1-preview/}, DeepSeek-R1-Distill-Qwen-7B\footnote{https://huggingface.co/deepseek-ai/DeepSeek-R1-Distill-Qwen-7B}, DeepSeek-R1-Distill-Qwen-32B\footnote{https://huggingface.co/deepseek-ai/DeepSeek-R1-Distill-Qwen-32B} and QwQ-32B~\cite{qwen2025qwen25technicalreport}. All the models use 
the prompt in Appendix~\ref{app:instrution} for reasoning. For each baseline, we only retain the prediction result after reasoning and use BM25 for further retrieval.

\paragraph{Query Reasoner with Reasoning-Intensive Retrievers (ReasonIR)} ReasonIR~\cite{shao2025reasonir} is the most recently acknowledged retriever specifically trained for reasoning-intensive retrieval tasks.
We further combine \ourmethod with ReasonIR for comparison to explore further improvements with the specialized reasoner and retriever in this task.

\subsubsection{Implementation Details}
With the initial checkpoint of Qwen2.5-7B-Instruct\footnote{https://huggingface.co/Qwen/Qwen2.5-7B-Instruct} and Qwen2.5-1.5B-Instruct\footnote{https://huggingface.co/Qwen/Qwen2.5-1.5B-Instruct}, \ourmethod 7B and 1.5B are both trained with TRL codebase\footnote{https://github.com/huggingface/trl} on a single node with 4 NVIDIA A800-80G GPUs. Following the instructions of Open-R1~\cite{openr1}, we use 1 GPU for vLLM~\cite{kwon2023vllm} serving and the rest 3 GPUs for model training. DeepSpeed~\cite{deepspeed}, ZeRO-3, and Gradient Checkpoint are applied to reduce the cost of VRAM. It takes about 16 hours for 1.5B model training and about 48 hours for 7B model training. We set the learning rate $1e-6$, the batch size per device $16$, and the KL coefficient $0.008$. For each input prompt, $16$ samples are generated to estimate the advantage in GRPO. Since we use bge-base-en-v1.5\footnote{https://huggingface.co/BAAI/bge-base-en-v1.5} embedding model to compute relevance, the maximum completion length is set to $500$ to avoid exceeding the input length limitation of the embedding model. 
Experiments on all the above-mentioned baselines  are conducted without reranking.

\subsection{Main Results}
Table~\ref{tab: bright_result} shows that our 7B model outperforms all query reasoning baselines of non-reasoning LLMs, including GPT-4o and DeepSeek V3,
performing comparable to the large reasoning models, \eg, o1-mini, QwQ-32B and DeepSeek R1. 
Our 7B model strikes a favorable balance between inference efficiency and reasoning performance, offering a compelling trade-off for query reasoning tasks. Besides, our 1.5B model  also achieves performance comparable to that of large-scale language models, making it an effective solution for resource-constrained scenarios.

To quantitatively assess the efficiency of different models, we report both their \textit{Performance} and \textit{Cost} in Table~\ref{tab:model_efficiency}. Here, \textbf{Performance} is defined as the nDCG@10 score achieved by each model on the BRIGHT benchmark~\cite{su2024bright} with BM25 retriever. Meanwhile, \textbf{Cost} represents the price of each model (USD per 1M output tokens) when accessed via the OpenRouter platform\footnote{https://openrouter.ai}, indicating the actual monetary expense required to obtain outputs from the model\footnote{We use the price of Qwen2.5-7B-Instruct as the price of \ourmethod-7B, and we define the price of \ourmethod-1.5B as $0.01$ since the price of Qwen2.5-1.5B is free.}. Based on the calculated efficiency (\textit{Eff} = \textit{Performance} / \textit{Cost}), \ourmethod-1.5B and \ourmethod-7B achieve the highest cost-effectiveness among all evaluated models, with efficiency scores of \textbf{2460.0} and \textbf{279.0}, respectively. This highlights the strong cost-performance advantage of the our method.
\begin{table}[htbp]
\centering
\small 
\begin{tabular}{lccc}
\toprule
\textbf{Model} & \textbf{Performance} & \textbf{Cost} & \textbf{Efficiency} \\
\midrule
GPT-4o & 26.5 & 10.0 & 2.7 \\
DeepSeek V3 & 27.8 & 0.9 & 31.6 \\
DeepSeek R1 & 29.6 & 2.2 & 13.6 \\
QwQ-32B & 28.9 & 0.2 & 144.5 \\
o1-preview & 30.2 & 60.0 & 0.5 \\
o1-mini & 28.3 & 4.4 & 6.4 \\
\ourmethod-7B & 27.9 & 0.1 & \underline{279.0} \\
\ourmethod-1.5B & 24.6 & 0.01 & \textbf{2460.0} \\
\bottomrule
\end{tabular}
\caption{Model performance (Perf.), cost, and efficiency (Eff. = Perf. / Cost).}
\label{tab:model_efficiency}
\end{table}

Compared to the retrievers specifically trained for reasoning, both the 7B and 1.5B models can both outperform ReasonIR with original query, and our 7B model can outperform Seed 1.5-Embedding. Since ReasonIR is an embedding model based on the backbone of LLaMA3.1-8B, the computational cost of pre-encoding documents can be prohibitive when the corpus is large. In contrast, \ourmethod can work with BM25 retrievers, incurring significantly lower pre-processing costs than LLM-based embedding models. 

As ReasonIR can work with reasoned queries to achieve better performance, we apply the reasoned queries generated by our method for further exploring the effect of combining reasoned queries with reasoning-intensive retrievers. We confirmed that our method can achieve further improvements based on ReasonIR. Our 1.5B model can outperform LLama3.1-8B-Instruct which is more than 5 times larger in parameters, and our 7B model can outperform GPT-4. Comparing with the reasoned queries of GPT-4 and our 7B model , the performance improvement based on ReasonIR ($29.9$->$31.9$) is higher than the improvement based on BM25 ($26.5$->$27.9$). These results incidate that our method is flexible and can work with different retrievers, the improvement of retriever will further expand the advantage of our method.
\subsection{Ablation Studies}
\begin{table*}[htbp]
\centering
\setlength{\tabcolsep}{2.8pt} 
\begin{tabularx}{\linewidth}{l|*{13}{>{\centering\arraybackslash}X}}
\toprule
\textbf{Dataset}  & Bio. & Earth. & Econ. & Psy. & Rob. & SO & SL & LC & Pony & AoPS & TQ & TT & \textbf{Avg}\\
\midrule
V1-QwQ   & 42.3 & 43.6 & 19.1 & \textbf{31.9} & 18.6 & \textbf{23.7} & 22.8 & 21.5 & 17.7 & 4.0 & 17.2 & 10.1  & 22.7 \\
V1-R1  & \textbf{48.0} & 46.5 & 19.9 & 31.4 & 15.2 & 23.6 & 22.3 & 21.2 & 18.0 & \textbf{5.3} & 18.8 & 11.3  & 23.5  \\
V2 & 46.0 & \textbf{47.1} & \textbf{21.1} & 31.2 & \textbf{19.8} & 21.7 & \textbf{24.3} & \textbf{22.5} & \textbf{21.8} & 4.3 & \textbf{19.7} & \textbf{15.9} & \textbf{24.6} \\
\bottomrule
\end{tabularx}
\caption{Results on different training data for \ourmethod-1.5B with BM25 retriever.}
\label{tab:dataset_performance}
\end{table*}
\subsubsection{Effect of Data Size and Quality}
\label{dataset_exp}
DeepSeek R1 performs better than QwQ-32B while the performance of V1-R1 and V1-QwQ is close on most BRIGHT subtasks. V1-R1 exhibits a notable advantage only in the Biology and Earth Science subtasks. We hypothesize that this may be attributed to the fact that these two subtasks are more knowledge-intensive compared to others, thereby granting the larger-parameter DeepSeek R1 model with 671B parameters a more pronounced advantage over QwQ-32B.
The V2 dataset with more samples leads to the best performance. Instead of using large reasoning models to generate answers for distillation, a better approach may be to use the answers selected by the users in the StackExchange datasets. It can be easily scaled since generating answers with large reasoning models on large-scale question set is too expensive. 

\subsubsection{Effect of Reinforcement Learning with Semi-Rule-Based Rewards}
\label{rl_sft}
We further explore the effect of our proposed approaches with semi-rule-based reward functions compared to traditional supervised fine-tuning (SFT). Following the same experimental settings in Section~\ref{dataset_exp}, we use Qwen2.5-1.5B-Instruct, with BM25 as retrievers. With the dataset of V1-QwQ and V2, we separately trained the model with SFT and RL. Results are shown in Table~\ref{tab:train_type}, where ``RL'' is for our proposed reinforced learning approaches and ``SFT'' is for supervised fine-tuning. Both RL and SFT are in full parameters.

These results indicate that when using the training data generated by large reasoning models, the performance of RL is slightly higher than SFT. - While using the user-selected answer data for training, the performance of SFT experienced a significant decline. This is likely because the user-selected answers written by actual users may exhibit substantial quality deficiencies (\eg, higher perplexity) compared to data synthesized by large reasoning models. In addition, we did not apply fine-grained data cleaning for the answer. As a result, the answers of the questions may include URL links of pictures which do not include available information. Using such data for supervised fine-tuning may lead to catastrophic forgetting in the model. In contrast, our proposed reinforcement learning approach with semi-rule-based reward functions does not strictly require the model to fit the answers per token exactly. Since the relevance score in reward function is based on the embedding similarity of generated answers and selected answers, the noisy signals in selected answers may not explicitly affect the similarity scores. As a result, our proposed approach demonstrates stronger generalization capabilities and greater tolerance for noisy data.
\begin{table}[htbp]
\centering
\begin{subtable}[t]{0.3\textwidth}
\centering
\begin{tabular}{p{1.2cm}p{1.2cm}p{0.8cm}}
\toprule
\textbf{Type} & \textbf{Resp. Length} & \textbf{Avg} \\
\midrule
Dense  & 500  & 23.5 \\
Sparse & 500  & 23.1 \\
Sparse & 1000 & 22.9 \\
\bottomrule
\end{tabular}
\caption{Relevance model types and Response length.}
\label{tab:reward_func}
\end{subtable}
\hfill 
\begin{subtable}[t]{0.3\textwidth}
\centering
\begin{tabular}{p{2.2cm}p{1.5cm}p{0.8cm}}
\toprule
\textbf{Content} & \textbf{Explicit} & \textbf{Avg} \\
\midrule
Answer              & No  & 22.7 \\
Think+Answer     & Yes & 21.4 \\
Answer              & Yes & 20.9 \\
\bottomrule
\end{tabular}
\caption{Effect of explicit thinking.}
\label{tab:think_explicit}
\end{subtable}
\hfill 
\begin{subtable}[t]{0.3\textwidth}
\centering
\begin{tabular}{p{1.5cm}p{1.5cm}p{0.8cm}}
\toprule
\textbf{Data} & \textbf{Method} & \textbf{Avg} \\
\midrule
V1-QwQ & SFT & 22.4 \\
V1-QwQ & RL  & 22.7 \\
V2     & SFT & 12.8 \\
V2     & RL  & 24.6 \\
\bottomrule
\end{tabular}
\caption{Training data and methods.}
\label{tab:train_type}
\end{subtable}
\caption{Performance comparison across (a) relevance model types and response length, (b) explicit thinking, and (c) training data and methods.}
\label{tab:all}
\end{table}

\subsubsection{Effect of Relevance Model in Reward Functions}
As we mentioned in Section~\ref{semi-rule-reward}, the relevance model is playing an important role in our proposed semi-rule-based reward functions. We further explore the effect of different relevance models in our proposed reward functions. Besides the dense embedding model of bge-base-en-v1.5, we also implement the relevance function via the sparse model of bge-m3~\cite{chen_bge_2024}. As the bge-m3 model can accept a longer input length, we also explore the effect of extending the maximum completion length to 1000. With Qwen2.5-1.5B-Instruct as base model and BM25 as retriever, we train the model on different reward functions and completion length settings on the training data of V1-R1. Results are shown in Table~\ref{tab:reward_func}.
Since all experiments are conducted with the sparse retriever of BM25, we initially expected bge-m3, as a sparse relevance model, to offer performance improvements. However, bge-m3 actually underperforms compared to the dense embedding model bge-base-en-v1.5, which has fewer parameters (110M vs 550M)\footnote{bge-m3 is based on XLM-RoBERTa-Large}. This result suggests that for our proposed semi-rule-based reward function, overly fine-grained relevance matching signals may harm the model’s generalization ability. It is worth noting that our training data is based on V1-R1 rather than V2, these results are unlikely to be primarily attributed to data noise since the answers are generated by DeepSeek R1. Furthermore, we observed that even increasing the output length did not improve performance, indicating that excessively long outputs might dilute the effective relevance signals, thus providing no benefit to final retrieval performance.
\subsubsection{Effect of Explicit Thinking}
Inspired by DeepSeek R1~\cite{guo2025deepseek} and some recent works~\cite{weller2025rank1,xie2025logicrl}, we further investigate the effect of the explicit thinking process. When the explicit thinking process is applied, the model will first think about the reasoning process explicitly and then provide the actual answer. The reasoning process and answer are enclosed with ``<think></think>'' and ``<answer></answer>'' tags. With the dataset of V1-QwQ, we train the model on Qwen2.5-1.5B-Instruct, and evaluate the query reasoners with BM25 retriever. Since the thinking process requires external output tokens, the max completion length is set to 1000 when the explicit thinking process is applied. Details about the prompt and reward settings are listed in Appendix~\ref{app:think_process}. Results are shown in Table~\ref{tab:think_explicit}. In the table, ``Explicitly Thinking'' denotes if the explicitly thinking process is applied for model training, and ``Content'' denotes if the output query contains the thinking process within the ``<think></think>'' tags. ``Thinking+Answer'' means that the contents within the ``<think></think>'' and ``<answer></answer>'' tags are concatenated as the reasoned queries, and ``Answer'' means that only the answer content is returned.

Experimental results indicate that applying an explicit thinking process does not improve performance on query reasoning tasks. Previous studies~\cite{weller2025rank1} have shown that explicitly generating the reasoning process within the ``<think></think>'' tags can be beneficial for certain reasoning-intensive tasks such as math or coding, possibly because these tasks require the model to produce answers in specific output formats. For example, in ranking tasks, the model receives a query and a document as input and must output a binary relevance judgment (true or false). In such cases, applying an explicit thinking process can help the model fully leverage its reasoning capabilities through chain-of-thought prompting, thereby enhancing inference performance. However, in the case of query reasoning tasks, the generated reasoned query inherently encapsulates the reasoning process and is not constrained by output format requirements. As a result, explicitly generating the reasoning process does not lead to further performance gains.
\section{Conclusion}
In this work, we present \ourmethod, a family of compact and efficient language models tailored for query reasoning and rewriting in reasoning-intensive retrieval. By leveraging the learning algorithm of GRPO with a novel semi-rule-based reward function, our approach enables effective and robust reinforcement learning without relying on expensive human-annotated datasets and retrieval sources. Our proposed models demonstrate strong performance on the BRIGHT benchmark, rivaling or even surpassing large-scale commercial LLMs, while significantly reducing inference cost and latency. Furthermore, \ourmethod models exhibit strong compatibility with both traditional and reasoning-intensive retrievers, making them highly versatile for real-world deployment. Our findings highlight a promising direction toward building lightweight, affordable, and high-performing reasoning components for retrieval-augmented generation pipelines and the latest deep research products.

\bibliography{main}
\bibliographystyle{icml2025}

\newpage
\appendix
\input{appendix}

\end{document}

%% file: appendix.tex
\onecolumn
\section{Prompt Templates}
\label{app:instrution}
\begin{figure*}[htbp]
\centering
\includegraphics[width=\linewidth]{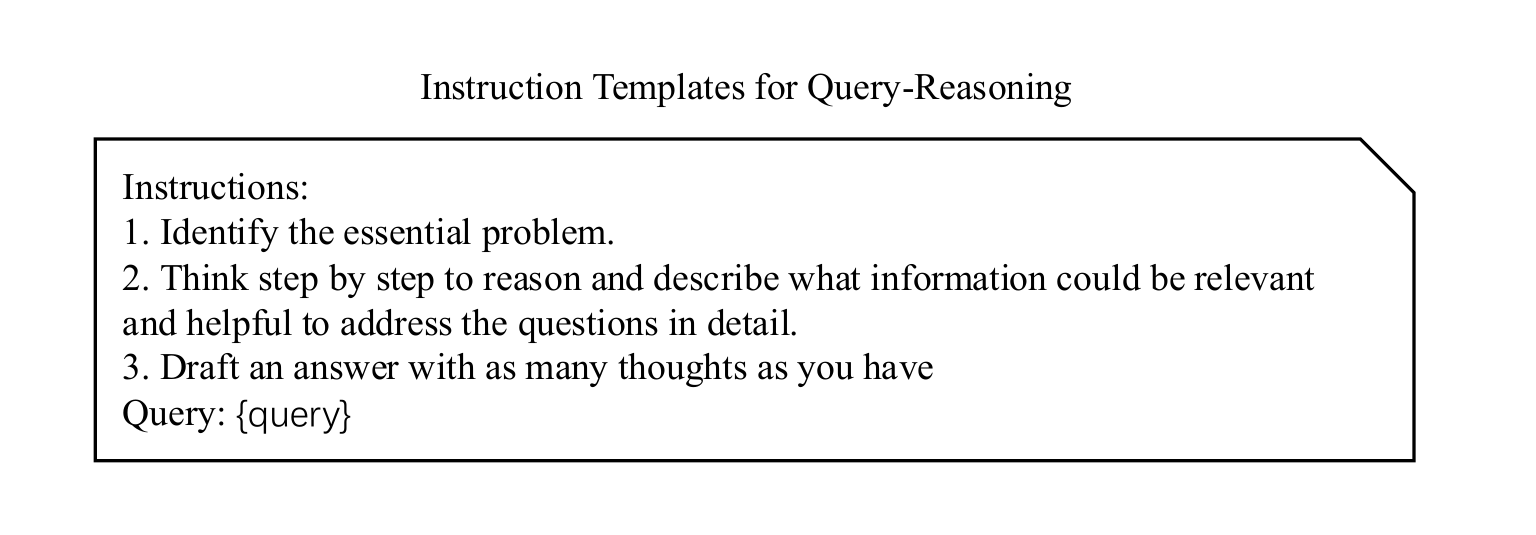}
\caption{
The prompt template for the instructions of Chain-of-Thought query reasoning.
}
\label{instruction_template_query_reasoning}
\end{figure*}

Figure~\ref{instruction_template_query_reasoning} shows the prompt template for the instructions of chain-of-thought query reasoning. The reasoner model takes the instructions and original query as input, and return a ``pseudo-answer'' with thoughts including as much relevant information as possible. The ``pseudo-answer'' can be used as the reasoned query, and the retriever can benefit from the external information provided by the reasoned query.
\section{Training Data}
\label{app:training_data_detail}
Details about the construction of training data are described as follows:
\begin{itemize}
    \item Version 1: sampling at most 1200 questions for each selected category to generate answers with large reasoning models. The selected categories include: 'biology', 'chemistry', 'codereview', 'cs', 'earthscience', 'economics', 'math', 'physics', 'robotics'. The Version 1 dataset includes around 10k sampled questions. In this paper, the corresponding datasets are denoted as \textbf{V1-R1} and \textbf{V1-QwQ}, indicating that the answers are generated by DeepSeek R1 or QwQ-32B.
    \item Version 2: sampling at most 1500 questions for each selected category with selected answers. Those questions can also be used to generate answers via large reasoning models. The categories include: 'ai', 'biology', 'chemistry','codereview', 'cs', 'earthscience', 'economics', 'computergraphics', 'math', 'mathoverflow', 'philosophy', 'physics', 'robotics', 'stackoverflow', 'sustainability', 'softwareengineering', 'bioinformatics’. The Version 2 dataset includes around 30k sampled questions, nearly three times as many as Version 1, making answer generation with large reasoning models unaffordable since the inference time is too long. In this paper, the dataset is denoted as \textbf{V2}.
\end{itemize}

\section{System Prompt and Reward for Explicit Thinking}
\label{app:think_process}
Inspired by previous works~\cite{xie2025logicrl,weller2025rank1}, we use the following system prompt to instruct the model to output the thinking process explicitly in the format of ``<think>thinking process</think><answer>the answer</answer>''. 
\begin{tcolorbox}[
  colback=cyan!5!white,
  colframe=teal!80!black,
  title=System Prompt,
  coltitle=white,
  colbacktitle=teal!80!black,
  fonttitle=\bfseries,
  boxrule=0.8pt,
  arc=2pt,
  left=2mm,
  right=2mm,
  top=1mm,
  bottom=1mm,
]

You are a helpful assistant. The assistant first thinks about the reasoning process in the mind and then provides the user with the answer. The reasoning process and answer are enclosed within \texttt{<think>} \texttt{</think>} and \texttt{<answer>} \texttt{</answer>} tags, respectively, i.e., \texttt{<think>} reasoning process here \texttt{</think>} \texttt{<answer>} answer here \texttt{</answer>}. 
\end{tcolorbox}
When the explicit thinking process is applied, we also design a format reward to force the model returning an output in the correct format. Our format checking strategy is identical to~\cite{xie2025logicrl}. If the model’s output fails the format checking, the reward function will immediately return a score of -1, and the subsequent computation of the query reasoning reward will be skipped. 

\section{License}
In this section we list the artifacts we used and the corresponding URL and licenses:

\begin{table}[htbp]
\centering
\renewcommand{\arraystretch}{1.2}
\begin{tabularx}{\textwidth}{l l X l}
\toprule
\textbf{Name} & \textbf{Type} & \textbf{URL} & \textbf{License} \\
\midrule
StackExchange-Preferences & Dataset & \url{https://huggingface.co/datasets/HuggingFaceH4/stack-exchange-preferences} & cc-by-sa-4.0 \\
BRIGHT Benchmark & Dataset & \url{https://huggingface.co/datasets/xlangai/BRIGHT} & cc-by-4.0 \\
Qwen2.5-1.5B-Instruct & Model & \url{https://huggingface.co/Qwen/Qwen2.5-1.5B-Instruct} & apache-2.0 \\
Qwen2.5-7B-Instruct & Model & \url{https://huggingface.co/Qwen/Qwen2.5-7B-Instruct} & apache-2.0 \\
bge-base-en-v1.5 & Model & \url{https://huggingface.co/BAAI/bge-base-en-v1.5} & mit \\
bge-m3 & Model & \url{https://huggingface.co/BAAI/bge-m3} & mit \\
QwQ-32B & Model & \url{https://huggingface.co/Qwen/QwQ-32B} & mit \\
DeepSeek R1 & Model & \url{https://huggingface.co/deepseek-ai/DeepSeek-R1} & mit \\
DeepSeek V3 & Model & \url{https://huggingface.co/deepseek-ai/DeepSeek-V3-0324} & mit \\
ReasonIR & Model & \url{https://huggingface.co/reasonir/ReasonIR-8B} & cc-by-nc-4.0 \\
\bottomrule
\end{tabularx}
\caption{List of datasets and models used, along with their URLs and licenses.}
\label{tab:resources}
\end{table}